# Engineering Effects of Vacuum Fluctuations on Two-dimensional Semiconductors

**Authors:** Jason Horng[1*], Yu-Hsun Chou[1,2], Tsu-chi Chang[2], Chu-Yuan Hsu[2], Tien-chang Lu[2], Hui Deng[1*]

**Affiliations:**

[1]Department of Physics, University of Michigan, Ann Arbor, Michigan 48109-1040, United States.

[2]Department of Photonics, College of Electrical and Computer Engineering, National Chiao Tung University, Hsinchu 300, Taiwan.

*Correspondence to: dengh@umich.edu, jahorng@umich.edu

**Abstract:**
The resonance energy and the transition rate of atoms, molecules and solids were understood as their intrinsic properties in classical electromagnetism. With the development of quantum electrodynamics, it is realized that these quantities are linked to the coupling of the transition dipole and the quantum vacuum. Such effects can be greatly amplified in macroscopic many-body systems from virtual photon exchange between dipoles, but are often masked by inhomogeneity and pure dephasing, especially in solids. Here, we observe an exceptionally large renormalization of exciton resonance and radiative decay rate in transition metal dichalcogenides monolayers due to interactions with the vacuum in both absorption and emission spectroscopy. Tuning the vacuum energy density near the monolayer, we demonstrate control of cooperative Lamb shift, radiative decay, and valley polarization as well as control of the charged exciton emission. Our work establishes a simple and robust experimental system for vacuum engineering of cooperative matter-light interactions.

The Lamb shift of the atomic transition frequency arises from the emission and re-absorption of virtual photons by single atoms.(*1*) Its discovery ushered the development of quantum electrodynamics and led to the surprising realization that the vacuum is not empty. The quantum fluctuations in vacuum couples with the transition dipole of matter and modifies its resonance energy and transition rate(*1–8*). The effects of the vacuum coupling become enhanced collectively in certain many-body systems by coherent exchange of virtual photons among the dipoles, leading to, for example, the cooperative Lamb shift(*5, 9, 10*) and superradiance(*6, 10–12*). However, strong dephasing and inhomogeneity in many-body systems have rendered these effects extremely difficult to observe experimentally. The cooperative Lamb shift in the optical domain has only been reported in a few experiments with cold atoms or ions(*9, 13–17*). In solids, it has only been observed in nuclei excited by synchrotron x-rays and in superconducting microwave circuits(*18, 19*).

Recently, high reflectivity was measured from a mere single layer of transition metal dichalcogenides crystal(TMDC)(*20, 21*), owing to the large exciton-photon coupling strength, near radiative-limited linewidth and two-dimensional(2D) translational symmetry. Based on this, theoretical work suggests monolayer TMDCs may provide an easy-to-access 2D many-body system for observing and utilizing the effects of vacuum fluctuations(*22, 23*). In this work, using a mirror to control the dipole transition of excitons in a high-quality, monolayer TMDC, we demonstrate the cooperative Lamb shift of the excitons accompanied by modified exciton radiative decay rate. This observation suggests that the TMDC monolayer can be used as a sensitive probe of the vacuum modes(*7, 24, 25*). We also demonstrate control of the charged exciton and valley polarization in the TMDCs with the tunable vacuum environment. In contrast to the strong coupling and Purcell effect, where a cavity is used to modify the resonances, in this work we have a fully open system coupled to a continuous spectrum of modes.

Figure 1A shows a schematic of our system. A monolayer TMDC is placed at a distance L from a mirror made of a distributed Bragg reflector(DBR) (Fig. 1A). The mirror imposes a boundary condition on the electromagnetic field in the space, creating a standing-wave pattern $E(r) = 2E_0 \sin(kr)$ with an electric field node at the mirror plane. Here $E_0$ and k are the electric field amplitude and the wavenumber of the incident planewave, and r is the distance from the mirror. This mode structure applies to not only classical fields but also the vacuum fluctuations. The structure of the vacuum fluctuations can be measured through its effect on a dipole emitter, such as an exciton in solids. Given a transition dipole, the radiative decay of the excited state to the ground state is proportional to the strength (spectral density) of electromagnetic fluctuations near the resonance frequency ω that are present in the environment. Therefore, measuring the lifetime of the excitation probes the local strength of vacuum fluctuations at the transition frequency. In effect, the monolayer can act as a local vacuum field analyzer.

The effects of vacuum fluctuations on the exciton resonance in a 2D material can be modeled by the following Hamiltonian in the rotating wave approximation:

$$H = \hbar\omega_0 b^\dagger b + \int dk\, \hbar\omega_k a_k^\dagger a_k - ig \int dk\, \sin(kL)\, (a_k b^\dagger + a_k^\dagger b) \quad (1)$$

where $b^\dagger, a_k^\dagger, b, a_k$ are creation and annihilation operators for an exciton and a photon, respectively, $\hbar\omega_0, \hbar\omega_k$ are the corresponding energies of the exciton and the photon, g is the dipole coupling constant between the exciton and photon field, L is the distance between the monolayer and the mirror. The factor $\sin(kL)$ represents the spatial mode structure of the electric field in front of a mirror. With the time-dependent Schrodinger equation, we solve for the exciton wavefunction X(t) and obtain the exciton resonance energy and radiative decay rate. As shown in the supplementary, the solution for $L \ll 2\pi c/\gamma$ ($\gamma$ is the radiative decay rate for a free-standing monolayer) can be written as:

$$X(t) = X(0)\exp[-i\widetilde{E_0}t/\hbar]\exp[-\frac{\tilde{\gamma}}{2}t] \quad (2)$$

where

$$\widetilde{E_0} = E_0 - \frac{\hbar\gamma}{2}\sin(2kL) \quad (3a)$$
$$\tilde{\gamma} = 2\gamma\cos^2(kL) \quad (3b)$$

The physical meaning of $\widetilde{E_0}$ and $\tilde{\gamma}$ is the renormalized exciton energy and exciton radiative rate, respectively, modified by the vacuum field.

The renormalized exciton radiative rate $\tilde{\gamma}$ equals $2\gamma$ a at the anti-node($kL = (n+0.5)\pi$) and zero at the node($kL = n\pi$) of the electric field. This is due to modification of exciton-photon coupling through local electric field. At the anti-node, the local field $2E_0$ enhances both the absorption and the emission rate, similar to the Purcell effect in a cavity configuration. At the node, local electric field is suppressed, therefore radiative decay of the exciton excitation is suppressed. The renormalized exciton energy $\widetilde{E_0}$ is shifted from $E_0$ due to coupling with the vacuum field. This energy shift in a collective excitation system has been named as "cooperative Lamb shift". It shares the same origin as the Lamb shift but can be as large as the radiative linewidth $\hbar\gamma$, due to the cooperative enhancement. In a radiative-limited sample, the renormalization on the resonance energy and the radiative decay rate can be directly observed through spectra of the exciton in frequency domain.

An alternative approach is to treat the renormalization as the result of interaction between the monolayer dipole moment, induced by vacuum fluctuations, and its image dipole(Fig. 1B). In our case, the image dipole has a $\pi$ phase difference compared with the original dipole due to the boundary condition set by the mirror. The renormalization of the radiative decay rate can be understood as follows: when the two dipole sheets are separate by $n\lambda$(the node condition), the radiated fields generated by them destructively interfere, leading to suppression in emission and apolariton of infinite lifetime(23). When their distance is $(n+0.5)\lambda$(the anti-node condition), the radiated fields constructively interfere leading to enhanced emission rate and short exciton lifetime. The cooperative Lamb shift can be understood through the dipole-dipole interaction: the emitted field from the image dipole at the original dipole has a phase shift of $\exp(i(\pi/2+2kL))$, where the $\pi/2$ comes from the phase relation between dipole and its radiation and $2kL$ is the phase accumulation during the propagation. Due to the dipole interaction $H=-\mu\cdot E$, the energy is lower if $L=[n/2, (n/2+0.25)]\lambda$ and the energy is higher if $L=[(n/2+0.25), (n/2+0.5)]\lambda$ leading the shift of exciton energy.

Observation of the cooperative Lamb shift and corresponding modulation of the radiative decay rate requires a homogeneous ensemble of emitters with nearly radiative-limited line broadening, which is challenging to realize in conventional semiconductors. In this work, using a monolayer TMDC, we observe the renormalization of both exciton energy $E_0$ and linewidth broadening $\hbar\gamma$ due to coupling with a radiation vacuum. To measure the renormalization of the exciton mode, we change the distance L between the monolayer and the mirror to modulate the local vacuum fluctuation at the 2D exciton position. We place an hBN-encapsulated $MoSe_2$ monolayer on a sapphire substrate in front of DBR mirror, whose position is controlled by a piezo-electric stage as illustrated in Fig. 1A. This setup allows *in situ* spectroscopy measurement of the same piece of $MoSe_2$ while we tune the local vacuum field.

First, we measure the reflection contrast spectra of the monolayer as it is moved through the standing wave of the electric field profile, using a weak femtosecond laser with bandwidth covering the exciton absorption peak. As shown in Figure2A, the absorption dip of the $MoSe_2$ A-exciton at1660meV is strongly modulated, following the period of the standing wave profile. Figure 2B shows several reflection contrast spectra (horizontal linecuts of Fig. 2A). The absorption dips are fit very well by Lorentzian functions(gray dashed curves), indicating minimal inhomogeneous broadening in the sample. The absorption depth is tuned from as low as 4% to as high as 99% at node and anti-node positions of the field, respectively. In the node region, we can hide the monolayer from the classical probe field even though it sits fully exposed in an open space. In the anti-node region, we can enhance the absorption to achieve the critical coupling condition where all photon energy is dumped into exciton energy, giving nearly 100% absorption.

We summarize in Fig. 2C the modulation of the absorption depth, linewidth and the resonance energy of the exciton as a function of mirror distance. We use the absorption depth to determine whether the monolayer is located at a node or anti-node(vertical gray dashed lines in Fig. 2C indicates the anti-nodal positions).In excellent agreement with predictions by Eq. (3), both the linewidth and energy of the exciton resonance show periodic modulations with a $\pi/2$ phase shift relative to each other and a modulation amplitude different by about a factor of 2.

The linewidth of the exciton changes by 2 meV, from 5.5meV at an anti-node to 3.5meV at a node (middle panel), corresponding to twice the non-renormalized radiative linewidth $\hbar\gamma$. Fitting the linewidth modulation with Eq. (3a)after including a constant offset$\gamma'$(blue dashed curve), we obtain $\hbar\gamma = 1.15 \pm 0.11\ meV$ and $\hbar\gamma' = 3.51 \pm 0.16 meV$. The offset of $\hbar\gamma'$accounts for contributions from other broadening mechanisms, including inhomogeneous, non-radiative and pure dephasing broadening. The$\hbar\gamma$ agrees with the radiative linewidth measured from linear reflection(*20*, *21*) and 2D spectroscopy(*26*, *27*). Similar linewidth modulations due to modified dipole-vacuum coupling has been observed in atomic systems(*13*, *14*) and superconducting qubits(*7*, *19*) in absorption spectra, but has not been demonstrated in a solid state system due to the small radiative linewidth compared with the total linewidth in typical semiconductor materials.

Cooperative Lamb shift of the exciton resonance due to exciton-vacuum coupling (lower panel of Fig. 2C) follows the same period of the linewidth modulation but $\pi/2$displaced in phase. The shift is zero at both the nodes and anti-nodes of the field, as predicted by Eq. (3b). Fitting the modulation of the exciton energy with Eq. (3b), we obtain the magnitude of the modulation $\hbar\gamma = 1.14 \pm 0.18 meV$, consistent with the result from the linewidth modulation.

Such cooperative Lamb shift has been predicted theoretically(*5*), but demonstrated only in atomic and superconducting qubit systems recently(*9,13,18,19*).Note that the observed linewidth and cooperative Lamb shift modification is on the order of 1meV(~250THz), which is much larger compared with atomic and superconducting qubits systems(tens of MHz or smaller),because of the extraordinary oscillator strength of the collectively coupled excitons.

Compared to absorption, incoherent photoluminescence (PL)stems from the coupling of incoherent exciton polarization to vacuum fluctuations, which cannot be described semi-classically. It is, therefore, interesting to test if the same renormalization effect appears in the emission spectrum of the exciton. Figure 3A shows a few examples of the PL spectra with varying L under the excitation of a continuous wave laser at 532nm. The features at 1662meV and 1632meV are due to exciton and trion emission, respectively. The linewidth of exciton emission measured at anti-nodes is clearly narrower than that measured at nodes(figure 3B), indicating that the same radiative decay rate renormalization is present in PL spectra. We summarize in Fig 3Cmodulation of the PL intensity, linewidth and resonance energy of the exciton peak. Reflection spectra are taken together to identify the node positions for exciton wavelength(750nm), which are labeled as vertical gray dashed lines in figure 3C. The PL intensity is complicated by the factor that the absorption of the 532nm pump laser is also modulated by the monolayer-mirror distance L. As a result, we observe PL intensity minima when the monolayer is located at the nodes of the field of either 532nm or 750nm, indicated by green arrows and gray dashed lines, respectively. The suppression of PL at the nodes of 750nm light arises from the optical interaction between the monolayer and its image dipole. To estimate the suppression/enhancement effect, we take the ratio of PL intensity at a node(step 16) versus an anti-node(step27) and compare the ratio to when no mirror is present. (Both points are around anti-node of the 532nm light.)In our measurements, the ratio can be as low as 5% and as high as 175%, showing control of PL quantum efficiency through radiative decay rate modulation.

The renormalization effect on the exciton decay rate can be observe in the frequency domain more directly. The third panel in figure 3B show the PL linewidth as a function of monolayer-mirror distance. The PL linewidth follows only the mode profile of the 750nm standing wave and not that of the 532nm excitation laser. It changes from about 4.5meV at the anti-nodes to about 2.2meV at the nodes. Correspondingly, a cooperative Lamb shift of 1.1meV is measured in PL(lower panel of Fig. 3B). Both results are in agreement with absorption measurements. Therefore we demonstrate the same exciton renormalization due to vacuum field in the emission domain, which is difficult to observe in atomic and superconducting qubit systems.

The dramatic effects of the vacuum on the TMDC excitons observed above demonstrate the possibility to control optical properties of TMDCs by vacuum engineering. We give two other examples, where we use the same tunable mirror approach to tune the charged exciton and valley polarization via the vacuum-matter coupling.

Charged exciton, or trions, are pronounced in TMDCs due to the strong Coulomb interaction in a two-dimensional film. Since the exciton and trion wavelengths are well separated, we can selectively enhance either the exciton or the trion emission by tuning the vacuum field strength at the respective wavelength. Figure 4Ashows two PL spectra from the same MoSe$_2$ monolayer, but at different mirror locations. The emission spectra show very different exciton/trion intensity ratio despite measured from the same position on the monolayer and at a fixed charge doping. This result shows that the ability to shape the vacuum-matter coupling allows us to control the interference effect for both exciton and trion emission. We show in Figure 4B the mirror-monolayer distance dependence of the PL intensities of exciton, trion and their ratio. The suppression of exciton emission is observed at the node for 750nm light indicated by the dashed gray lines while the suppression of trion emission is observed at the node for 780nm light indicated by the orange arrows. The ratio of exciton and trion emission intensity changes over two orders of magnitude, from 0.02 to 2.48. This simple technique can be utilized to various 2D materials applications to enhance or suppress the transition of interest.

Another important property of TMDC materials is the valley degree of freedom. However, the valley polarization of the excitons is often complicated by several competing decay and exchange mechanisms. Considering the two dominant mechanisms -- the radiative decay and inter-valley exchange interaction -- we can formulate the valley polarization, quantified by the degree of circular polarization (DOCP) as

$$\text{DOCP} = \frac{I_+ - I_-}{I_+ + I_-} = \frac{\tilde{\gamma}}{\tilde{\gamma} + 2\eta} \quad (4)$$

where $\tilde{\gamma}$ and $\eta$ are the radiative decay rate and intervalley exchange rate, respectively(*28*). For this study, we use an encapsulated WSe$_2$ sample and 633nm laser with σ+ polarization for excitation. The exciton PL spectra around 1.74eV with anti-node, node and no mirror condition are shown in Fig. 4C. The measured DOCP of exciton as a function of mirror distance(Fig. 4D) shows oscillation behavior in accord with the reflection contrast modulation. The tuning of vacuum fluctuation through mirror distance allows us to change the radiative decay rate γ of the WSe$_2$ exciton, resulting in enhancement (suppression) of DOCP at anti-nodes (nodes). Given that the measured DOCP without mirror present is 37%(blue dashed line in Fig. 4D), the modulation should be from 0% to 54% based on Eq. (4) when the radiative decay rate is tuned from 0 to 2γ. However, we observe a modulation from 25% to 43% experimentally. The discrepancy might come from other depolarization mechanisms or effective radiative lifetime from dark excitons in the WSe$_2$ monolayer and require further studies. Nonetheless, the modulation via simply a mirror is already comparable with other reports using microcavities(*28–30*).

In conclusion, we observe the renormalization of exciton resonance energy and radiative linewidth in TMDC monolayers due to coupling to vacuum fields. This effect has only been observed in atomic and superconducting qubit systems before. The tightly bound exciton in TMDC monolayers leads to an exceptionally large radiative linewidth, enabling pronounced effects of vacuum engineering, manifested as a large cooperative Lamb shift, strong modulation of the radiative linewidth, and control over trion and valley degrees of freedom. Our study shows intriguing collective physics of vacuum effects on excitonic many-body systems and could pave the way for future quantum optics research with 2D materials.

Note: Upon the completion of this manuscript, we became aware of preprints of related works by You Zhou, et. al.(*31*) and Christopher Rogers, et. al.(*32*).

**Acknowledgments:** We gratefully thank Professor Mack Kira for fruitful discussions.

**Funding:** This work was supported by the Multidisciplinary University Research Initiative(MURI) program by the Army Research Office(ARO) under Grant # W911NF-17-1-0312. Y.-H. C. acknowledges the support from Graduate Student Study Abroad Program(GSSAP) from Ministry of Science and Technology in Taiwan. This work was partially supported by the Minister of Science and Technology (Taiwan) under contracts Nos. 106-2917-I-564-021.

**Author contributions:** J.H. and H.D. conceived the research. J.H. designed the experiments, carried out optical measurements, analyzed the data and performed theoretical analysis. T.-C. C., C.-Y. H. and T.-C. L. fabricated DBR mirrors. Y.-H. C. fabricated the TMDC samples assisted by J.H.. J.H. and H.D. wrote the manuscript with inputs from all authors.

**Competing interests:** Authors declare no competing interests.

**Data and materials availability:** All data is available in the main text or the supplementary materials.

## Figures

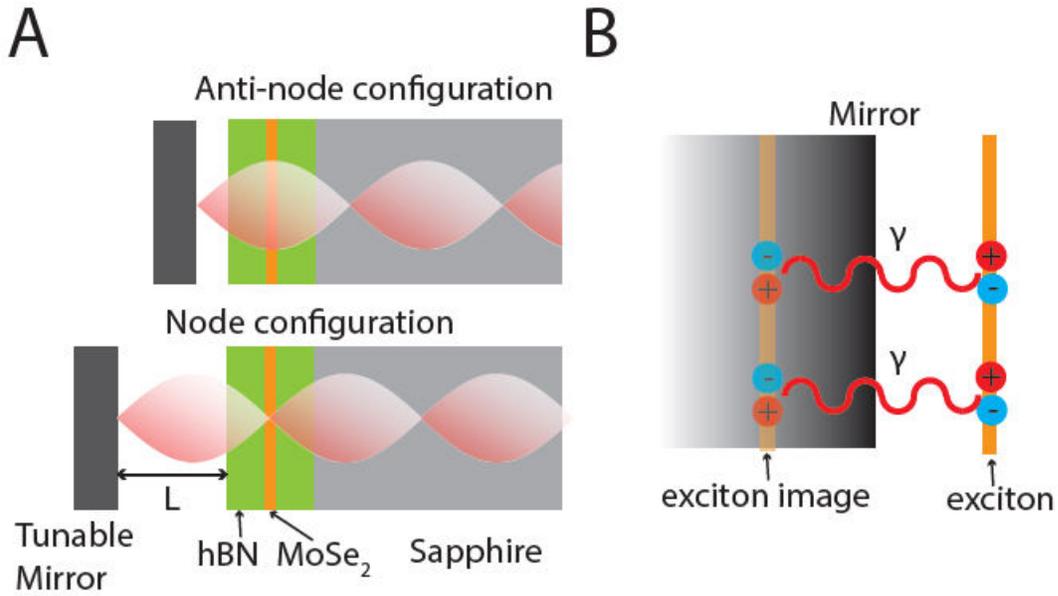

**Fig. 1. Modifying vacuum fluctuations at a two-dimensional semiconductor in front of a mirror.** (**A**) A monolayer MoSe$_2$ is placed in front of a mirror with a tunable distance L. Depending on the mirror distance L, the monolayer samples different vacuum fluctuation due to the standing wave imposed by the mirror boundary condition. Altering the vacuum-monolayer coupling leads to renormalization of exciton resonance energy and radiative decay rate. (**B**) Another approach to understand the system is to consider the exciton in the MoSe$_2$ monolayer interacting with its mirror image through dipole-dipole interactions. Due to the macroscopic dipole moment from two-dimensional excitons, the renormalization effect can be significant.

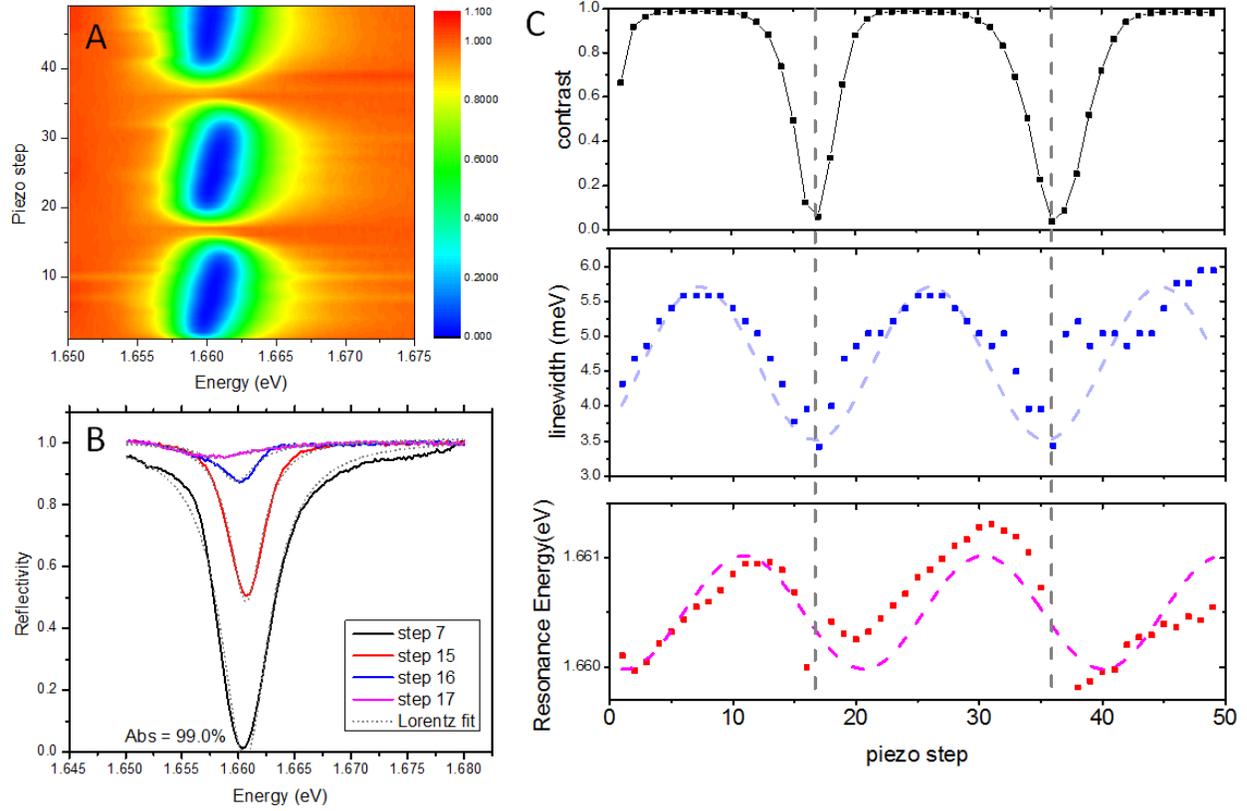

**Fig. 2. Effects of vacuum fluctuations on the exciton transition measured via absorption spectroscopy.** (**A**) Measured reflection contrast of a MoSe$_2$ monolayer in front of a distributed-Bragg-reflector as a function of photon energy and monolayer-mirror distance L. The absorption dip around 1660meV corresponds to the A-exciton resonance. (**B**) Several spectra from (A) showing the shift and broadening of the exciton absorption when the monolayer is moved from anode (step 7) to an anti-node(step 17) of the field. (**C**) Mirror-position dependence of the depth (top panel), linewidth (middle panel) and resonance energy (bottom panel) of the A-exciton absorption dip. The anti-node positions are identified by the maximum absorption depth (~99%), while the node positions are identified by the minimum absorption depth (~0.04%) and marked by the dashed lines. The modulation of the vacuum fluctuations lead to modulations of both the linewidth and the cooperative Lamb shift, which are fit but sinusoidal functions with a $\pi/2$ relative phase shift (blue and red dashed curves, respectively).

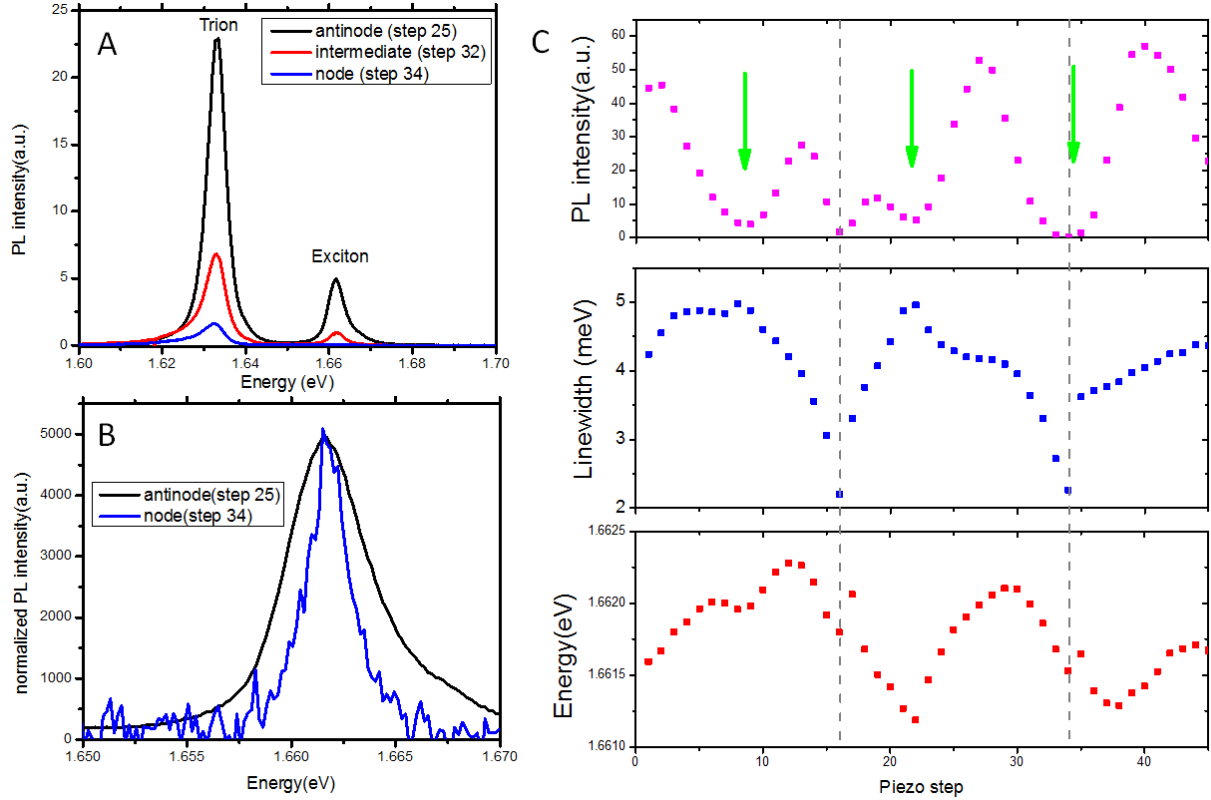

**Fig. 3. Effects of vacuum fluctuations on the photoluminescence(PL) properties of a MoSe$_2$ monolayer.** (**A**) Measured PL spectra of a MoSe$_2$ monolayer in front of a mirror as the monolayer is moved from an anti-node (black line) to a node (blue line) of the modified vacuum field. The emission peaks around 1660 and 1630 meV correspond to the A-exciton and trion resonances, respectively. (**B**) Normalized PL spectra at an anti-node and a node, showing different linewidths. (**C**) Mirror-position dependence of the intensity, linewidth and resonance energy of the A-exciton PL, showing modulations following the modified vacuum fluctuations. The PL resonance energy also shows the cooperative Lamb shift. The vertical dashed lines mark the nodes of the vacuum field identified from absorption spectra. The green arrows indicate where the absorption of the 532 nm excitation laser is suppressed.

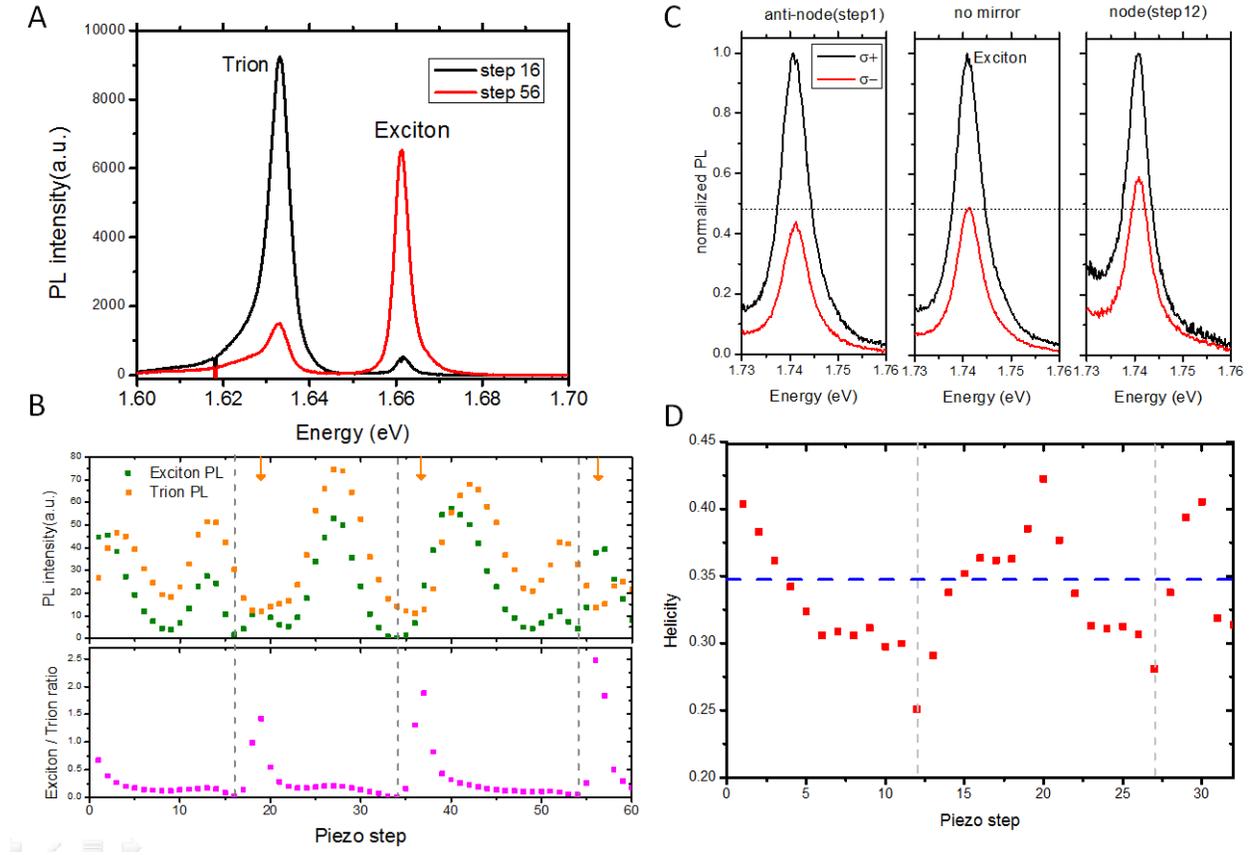

**Fig. 4. Controlling the emission properties of 2D materials via vacuum fluctuations.** (**A**) Two MoSe$_2$ emission spectra measured at the same position on the monolayer with a fixed doping. (**B**) The exciton and trion emission intensities (top) and their ratio (bottom) as a function of the monolayer-mirror distance, showing enhancement and suppression of the exciton relative to the trion emission with varying distances. (**C**) Helicity-resolved PL spectra of monolayer WSe$_2$ at the field anti-node (left) and node (right) in front of a mirror, and without a mirror (middle). (**D**) Degree of circular polarization(DOCP) vs. the mirror position. It changes from 25% to 40%, showing the effect of vacuum coupling on the valley dynamics of TMDCs. The blue dashed line indicate the DOCP when mirror is no present.